\documentclass[prb,aps,twocolumn,floatfix,showpacs]{revtex4}
\usepackage[pdftex]{graphicx}
\usepackage{bm}
\usepackage{color}
\usepackage{amssymb}
\usepackage{mathrsfs}
\usepackage{amsmath}
\usepackage{undertilde}

\newcommand{\etal}{{\it et al.}\hspace{1mm}}
\newcommand{\newc}{\newcommand}

\newc{\be}{\begin{equation}}
\newc{\ee}{\end{equation}}
\newc{\bfe}{\begin{floatequation}}
\newc{\efe}{\end{floatequation}}
\newc{\bea}{\begin{eqnarray}}
\newc{\eea}{\end{eqnarray}}
\newc{\ie}{{\it i.e.} }
\newc{\eg}{{\it e.g.} }
\newc{\etc}{{\it etc.} }
\newc{\ra}{\rightarrow}
\newc{\lra}{\leftrightarrow}
\newc{\lsim}{\buildrel\langle\over{\sim}}
\newc{\gsim}{\buildrel\rangle\over{\sim}}

\begin{document}
\title{The Karlqvist approximation revisited}
\author{C. Tannous}
\affiliation{Laboratoire de Magn\'etisme de Bretagne - CNRS FRE 3117\\
UBO, 6, Avenue le Gorgeu C.S.93837 - 29238 Brest Cedex 3 - FRANCE}

\date{\today}

\begin{abstract}
The Karlqvist approximation signaling the historical beginning of 
magnetic recording head theory is reviewed and compared to 
various approaches progressing from Green, Fourier, Conformal
mapping that obeys the Sommerfeld edge condition at angular points and 
leads to exact results.
\end{abstract}
\maketitle

\section{Introduction}
Karlqvist seminal historical paper~\cite{Karlqvist} is an important
landmark and the first building block
of magnetic recording theory. Despite all ensuing developments in
recording theory it is still used as a standard reference
with respect to any model aiming at the description of recording
or replay head fields.

Karlqvist work in based on magnetostatics and Green's function theory
of partial differential equations since the basic description of
head field physics is based on Laplace or Poisson equations 
the magnetostatic potential must satisfy under certain conditions.

Magnetic playback theory deals with time dependence since tapes and hard 
disk platters move with respect to head in contrast with 
our work that deals with static head description. 

We concentrate on the description of magnetostatic interactions
in 3D simplifying them to 2D in order to fully understand the Karlqvist
approximation and its many aspects that lie at the roots of recording theory.

This work can be taught as an application chapter 
in a general Electromagnetism course at the Graduate level 
or in an undergraduate class of Partial Differential Equations 
since physicists might be interested in applying some mathematical
methods to solve Laplace and Poisson equations or understanding issues 
surrounding the development of magnetic recording advances in 
hard disks and storage media.

\begin{figure}
  \begin{center}
    \begin{tabular}{c}
   \includegraphics[angle=0,height=60mm,clip=]{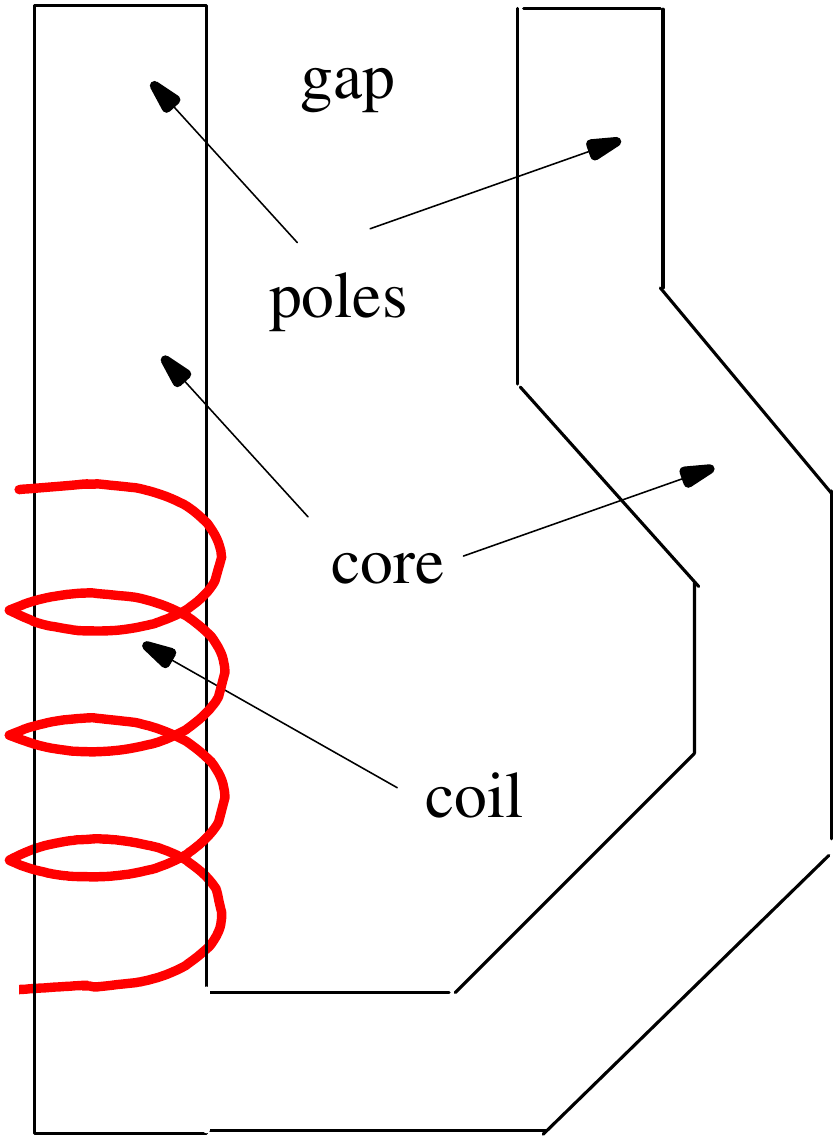} \\
          (a)  \\
   \includegraphics[angle=0,height=60mm,clip=]{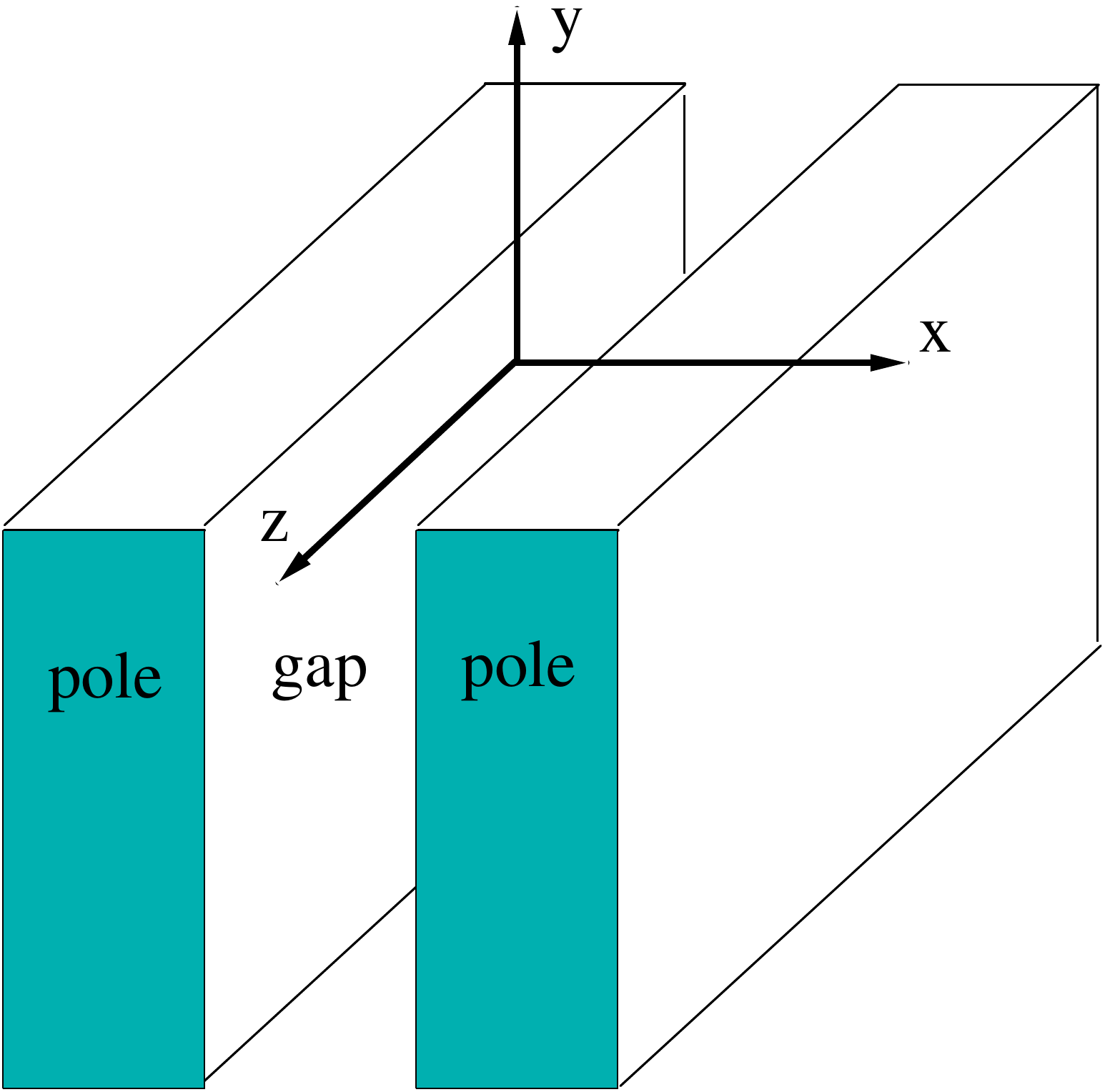} \\
           (b) \\
       \end{tabular}
    \caption{(a) General structure of a magnetic recording head. Pole and core 
are generally made from very soft material possessing a very large permeability $\mu$. 
They are excited by a coil that will create a magnetic field in the gap and around the pole surfaces.
(b) 3D perspective view of a simplified magnetic recording head tip. }
    \label{fig1}
  \end{center}
\end{figure}

Karlqvist approximation is based on a simplified description of the recording
head physics (see fig.~\ref{fig1} and fig.~\ref{fig2}) to 2D  such that it consists of two 
semi-infinite poles made from a very soft material possessing
an infinite permeability. The poles are separated by a gap region and
face a semi-infinite plane devoid of any magnetic charges.

In addition a number of simplifying assumptions are made in order to 
obtain a fully analytic solution to the field created by the poles, such as:

\begin{enumerate}
\item Uniform magnetization above poles leading to a zero volume source density.
\item Spatially linear surface source density in the gap.
\item Uniform field $H_g$ in the gap all the way up to $y=0$. In reality, it
can be uniform only in the case $y<<0$ far from the $\partial \Omega$ boundary.
\item Ignorance of Sommerfeld radiation condition~\cite{Green} near angular boundaries such as 
regions around points $x=\pm a,y=0$.
\end{enumerate}

In this work we review briefly how magnetostatic theory is used for the
description of fields created by the head and its interaction with recording
media. It is organised as follows. Section 2 describes Green's function theory
in 3D and its application to 2D with the Karlqvist approximation
delivering the potential and field created by the recording head in the plane
on the basis of a simplified Boundary Value Problem (BVP) described above.
In section 3, Karlqvist solution is revised and compared to exact Fourier methods 
meant to solve the exact BVP.
In section 4 we use conformal mapping methods to retrieve the solution for a more
realistic head accounting for edge condition at angular points and 
leading to exact results. Finally section 5 bears conclusions and 
perspectives of the work.

\section{Green's function approach}

In the 3D magnetostatic approximation, Maxwell equations for fields 
$\bm{E}$ and $\bm{H}$ reduce to:

\be
\bm{\nabla} \times \bm{E}=0,  \hspace{2cm} \bm{\nabla} \times \bm{H}=0
\ee

In the absence of magnetization and poles ($\bm{\nabla} \cdot \bm{B}=0$)
we infer existence of a magnetic scalar potential $\phi(\bm{r})$ at
any point $\bm{r}$ leading to $\bm{H}$ definition and Laplace equation:
\be
\bm{H}=-\bm{\nabla} \phi(\bm{r}), \hspace{2cm}  \Delta \phi(\bm{r})=0 
\label{Laplace} 
\ee

In presence of a magnetization field $\bm{M}(r)$, the lowest approximation
for the "far-field"  potential $\phi(\bm{r})$ in a multipole expansion is dipolar 
leading to a volume source $\bm{\nabla} \cdot \bm{M}$ and 
a surface density source $\bm{n} \cdot \bm{M}$ where $ \bm{n}$  is the outward normal 
to the finite surface sample embodying  $\bm{M}(r)$.

The potential $\phi(\bm{r})$ can be obtained, from volume source density
$\rho(\bm{r})=-\bm{\nabla} \cdot \bm{M(r)}$ and surface source density
$\sigma(\bm{r})=\bm{n} \cdot \bm{M(r)}$ as (considering free-space permeability $\mu_0=1$):

\be
\phi(\bm{r})=\frac{1}{4\pi} \int_V \frac{\rho(\bm{r'}) dV'}{|\bm {r -r'}|}
+ \frac{1}{4\pi} \oint_{\partial V} \frac{\sigma(\bm{r'}) dS'}{|\bm {r -r'}|}
\label{potential}
\ee

This result can in fact be reinterpreted as a BVP with a
Poisson equation subjected to presence of volume source and surface sources:

\be
\Delta \phi(\bm{r})=\bm{\nabla} \cdot \bm{M}, \hspace{2mm} \mbox{in} \hspace{2mm} V, 
\hspace{5mm} \phi(\bm{r})=\bm{n} \cdot \bm{M} \hspace{2mm} \mbox{on} \hspace{2mm} \partial V
\label{BVP3}
\ee 

It is interesting to note that the surface density is equivalent to a surface 
potential $\phi_S(\bm{r})= \bm{n} \cdot \bm{M(r)}$. 

The BVP eq.~\ref{BVP3} can be solved with Green second identity:

\be
\int_V [\phi(\bm{r}) \Delta G - G  \Delta \phi(\bm{r})]  dV =  \oint_{\partial V} 
[\phi(\bm{r})  \bm {\nabla} G - G  \bm {\nabla} \phi(\bm{r})] \cdot \bm {n} dS
\ee

derived from coupling Laplace equation and the definition of the Green's function solution of:

\be
\Delta G(\bm {r,r'})=\delta (\bm {r-r'})
\label{Green}
\ee

where $\delta (\bm {r})$ is Dirac delta function.

The radial Green's function expression is obtained by solving the 3D Laplacian in spherical
coordinates:

\be
\frac{1}{r^2} \frac{\partial}{\partial r} \left( r^2 \frac{ \partial G}{\partial r} \right)=0 
\hspace{2mm} \mbox{for} \hspace{2mm} r> 0
\ee

Thus the full Green function $G(\bm {r,r'})=-\frac{1}{4\pi} \frac{1}{|\bm {r -r'}|}$ 
is obtained (the angular factor $4\pi$ originates from solid angle subtending all 3D space) and
exploited in expressing the potential as:

\be
\phi(\bm{r})= \int_V G(\bm {r,r'}) \rho(\bm{r'}) dV'
 + \oint_{\partial V} \bm {\nabla'} G(\bm {r,r'}) \cdot \bm {n'} \phi_S(\bm{r'}) dS'
\label{Green3D}
\ee

Using eq.~\ref{Green}, eq.~\ref{BVP3} and the condition $ \bm {\nabla} \phi(\bm{r})=0 $ on $\partial V$
we retrieve the general solution for the potential eq.~\ref{potential}.

\subsection{Karlqvist 2D magnetostatic solution}

Karlqvist approximation is based on approximating the magnetic recording head
as consisting of the tips only (see fig.\ref{fig1}) with the assumption
the poles are large and thick and made with an infinitely soft material (permeability
$\mu \approx \infty$).

\begin{figure}[htbp]
  \centering
    \includegraphics[angle=0,width=80mm,clip=]{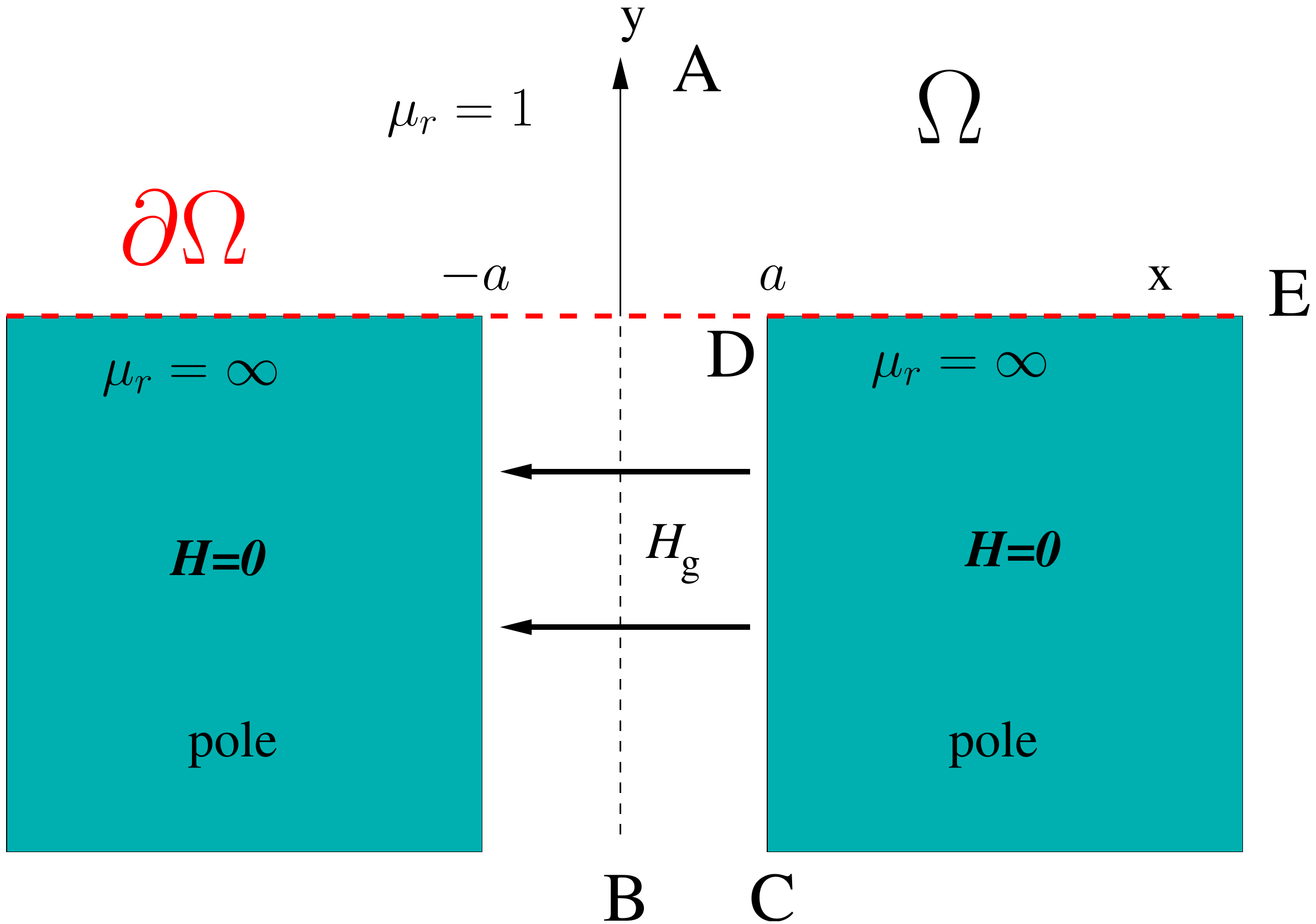}  
\caption{(Color on-line) 2D approximation of a magnetic recording head. 
BVP1 eq.~\ref{BVP1} is defined by domain $\Omega$ ($y>0$) bounded by $\partial \Omega$ (in dashed red), 
the entire $x$ axis. The poles are made from very soft material, $H=0$ for any point on 
$\partial \Omega$ and inside poles since $H_g$ is finite in the gap region. 
Note vertices $x=\pm a,y=0$ that should be subject to Sommerfeld radiation condition. }
\label{fig2}
\end{figure}

A flat 2D model for the recording head displayed in fig.~\ref{fig2} can be made
since poles are considered as infinitely large (along $x$ direction as in fig.~\ref{fig1}) 
and thick (along $z$ direction, see fig.~\ref{fig1}) separated by the gap region.

With this assumption, one cannot use the potential result previously 
given in eq.~\ref{potential} since this is valid strictly in the 3D case.

Thus we move to a fully 2D approach to solve the BVP posed by the magnetic 
head stated mathematically as:

\be
\Delta \phi(\bm{r})=0 \hspace{2mm} \mbox{in} \hspace{2mm} \Omega, 
\hspace{5mm} \phi(\bm{r})=\phi_S(\bm{r}) \hspace{2mm} \mbox{on} \hspace{2mm} \partial \Omega \hspace{2mm} \mbox{(BVP1)}
\label{BVP1}
\ee

This means the magnetization field is uniform yielding zero volume source, however we have 
a non-zero surface potential $\phi_S(\bm{r})$ on $\partial \Omega$.

The Green's function of the 2D Laplace equation
is $ G(\bm {r,r'})=\frac{1}{2\pi} \ln|\bm{r-r'}|$ obtained by solving the 2D Laplace equation
in polar coordinates for the radial part (as in the 3D case, $2\pi$ corresponds to angle subtending the entire plane):

\be
\frac{\partial^2 G}{\partial r^2} + \frac{1}{r} \frac{\partial G}{\partial r}=0 
\hspace{2mm} \mbox{for} \hspace{2mm} r> 0
\ee

As in the 3D case, the solution to the BVP is obtained from Green second identity
over the domains $\Omega$ and $\partial \Omega$ corresponding to the
previous 3D volume and surface $V$ and $\partial V$ respectively using an adequate 2D Green function.

The 2D Green function $ G(\bm {r,r'})=\frac{1}{2\pi} \ln|\bm{r-r'}|$ is for the entire plane whereas in our case, the domain $\Omega$ is limited to the upper half-plane, thus
the Green function should be  $ G(\bm {r,r'})=\frac{1}{2\pi} \ln|\bm{r-r'}|- 
\frac{1}{2\pi} \ln|\bm{r-\utilde{r'}}| $ using
the method of images~\cite{Green}.  $\bm{\utilde{r'}}=(x',-y')$ is the image of $\bm{r'}=(x',y')$
in the lower half-plane.

The explicit form of the Green function in the upper half-plane becomes:
\bea
G(\bm {r,r'})&=&\frac{1}{2\pi}\ln \left[ \frac{\sqrt{{(x-x')}^2+{(y-y')}^2}}{\sqrt{{(x-x')}^2+{(y+y')}^2}} \right] \nonumber \\
&=&\frac{1}{4\pi}\ln \left[ \frac{{(x-x')}^2+{(y-y')}^2}{{(x-x')}^2+{(y+y')}^2} \right]
\eea

Using eq.~\ref{Green}, eq.~\ref{BVP1} and $ \bm {\nabla} \phi(\bm{r})=0$ on $\partial \Omega$:
\be
\phi(\bm{r})= \int_{\partial \Omega} \phi_S(\bm{r'}) \bm {\nabla'} G(\bm {r,r'}) \cdot \bm{n'} dS'
\label{2dgreen}
\ee

The quantity $\bm {\nabla'} G \cdot \bm{n'}$ over the boundary $\partial \Omega$ is obtained with the
outward normal $\bm{n'}=(0,-1)$ as:

\be
\bm {\nabla'} G \cdot \bm{n'}=-\left.{\frac{\partial G(\bm {r,r'})}{\partial y'}}\right|_{y'=0}=
\frac{y}{\pi} \frac{1}{[{(x-x')}^2+y^2]}
\ee

yielding: 

\be
\phi(\bm{r})= \phi(x,y)=  \frac{y}{\pi} \int_{-\infty}^{+\infty}
 \frac{\phi_S(x') dx'}{{(x-x')}^2+y^2}
\label{2dgreen1}
\ee

The above result called the Poisson integral for the half-plane may be derived
from the fact an analytic function can be obtained at any point from the values it takes 
over a boundary (see for instance Ablowitz \etal~\cite{Ablowitz}). It can be viewed as
a convolution between the source function $\phi_S(\bm{r}) $ and the Green function.
The result  is retrieved with Fourier analysis in the next section.

Along the boundary $\partial \Omega$  the source potential function $\phi_S(x)$ produces 
the field $H_g$ in the gap interval $[-a,+a]$ thus $H_g=-\frac{d\phi_S}{dx}$ using the
definition eq.~(\ref{Laplace}). Outside
$[-a,+a]$ interval, the potential is constant and the field is zero.\\

In fact the field is created by a current injected in the coil (see fig.~\ref{fig1}),
as given by Amp\`ere law: $NI=H_g$ with $I$ the current and $N$ the number of coil
turns.
 
The resulting source potential over $\partial \Omega$   is given by: \\

\be
\phi_S(x)= \left\{
\begin{array}{rlc}
 -V_0 &   &  x< -a, \\
 H_g x  &   & -a<x< a, \\
 V_0   &   & x> a  
\end{array}
\right. 
\ee

Since $\phi_S(x)$ is a continuous function we infer that $V_0=H_g a=NIa$.

Over the region $\Omega$, the evaluation of the potential $\phi(x,y)$  requires integration
of eq.~\ref{2dgreen} to get:

\begin{widetext}
\be
\phi(x,y)= \frac{H_g}{\pi} \left[ (x+a) \tan^{-1} \left( \frac{x+a}{y} \right) 
-(x-a) \tan^{-1} \left( \frac{x-a}{y} \right)
- \frac{y}{2} \ln \frac{{(x+a)}^2+y^2}{{(x-a)}^2+y^2} \right]
\label{2dpot}
\ee
\end{widetext}

\begin{figure}[htbp]
  \centering
    \includegraphics[angle=0,width=85mm,clip=]{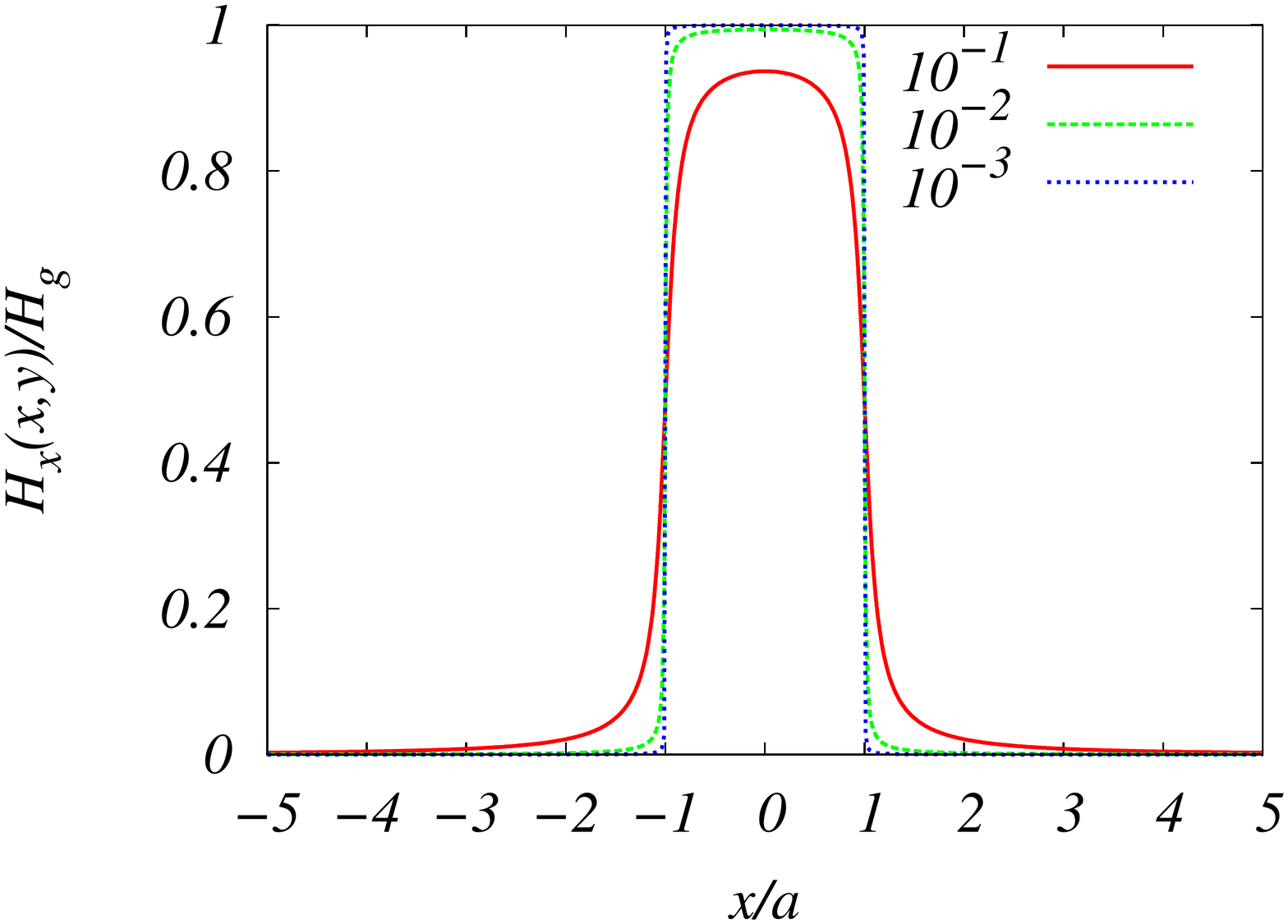} 
   \includegraphics[angle=0,width=85mm,clip=]{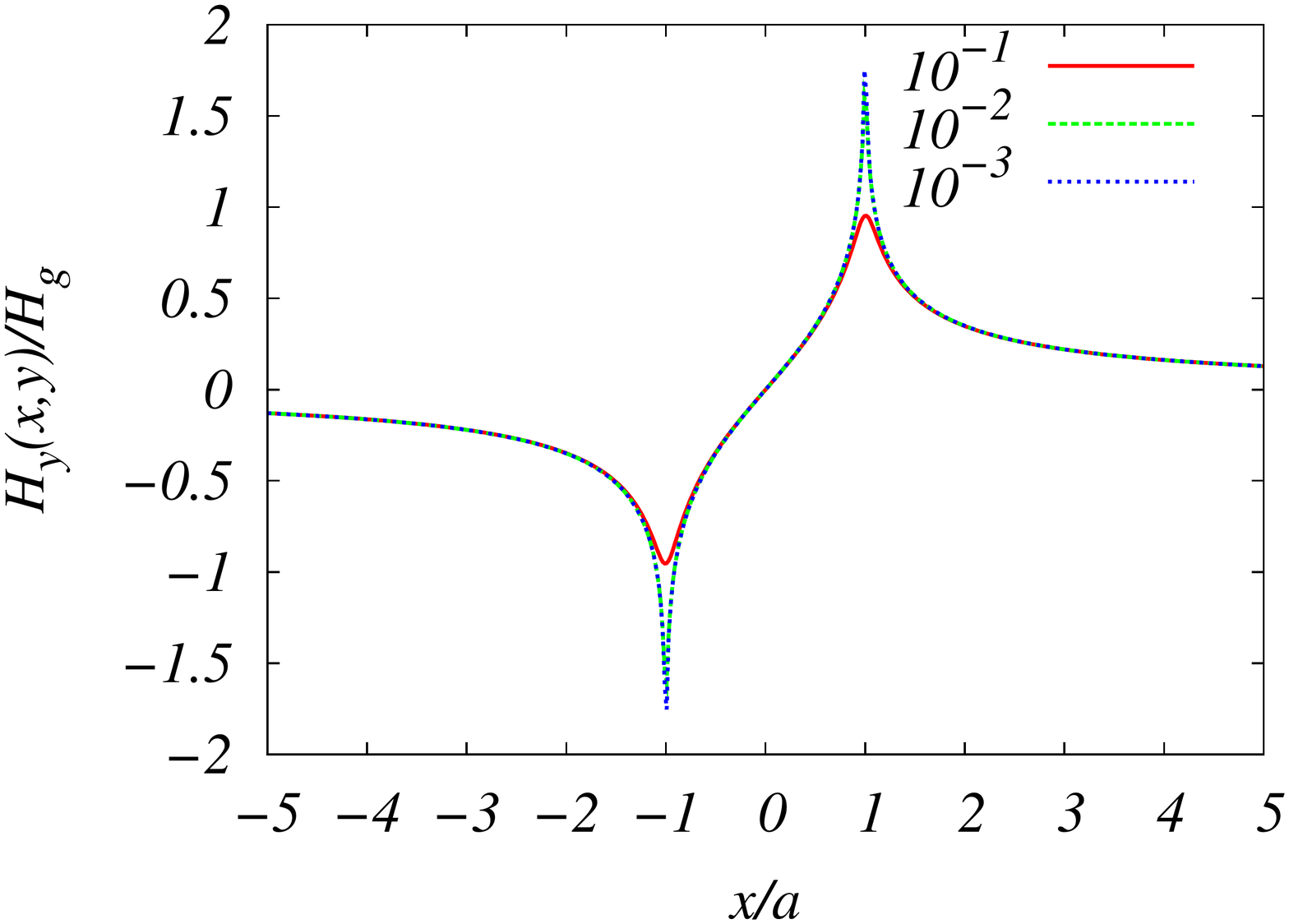}  
\caption{Variation of the fields $H_x(x,y), H_y(x,y)$ according to the Karlqvist approximation with $x$ for fixed $y/a=10^{-1},10^{-2},10^{-3}$. The edge effect is not visible in $H_x(x,y)$ whereas it is weaker than reality for $H_y(x,y)$ (see Conformal Mapping section).}
    \label{Karlqvist}
\end{figure}

The field components $H_x(x,y), H_y(x,y)$ over $\Omega$ obtained from definition~(\ref{Laplace})
are drawn versus $x/a$ for a fixed value of $y/a$ in fig.~\ref{Karlqvist}.

\subsection{Fourier Transform solution}

We retrieve the Green function solution of the Boundary-value problem eq.~\ref{BVP1} 
by introducing the Fourier transform:

\be
\Phi(k,y)={\cal F}[\phi(x,y)]=  \frac{1}{2\pi} \int_{-\infty}^{+\infty} \phi(x,y) e^{ikx} dx
\label{FT}
\ee

Laplace equation is transformed into:

\be
\frac{\partial^2 \Phi(k,y)}{\partial y^2}-k^2 \Phi(k,y)=0 \hspace{2mm} \mbox{for} \hspace{2mm} y> 0
\ee

The solution $\Phi(k,y)$ being a superposition of $e^{|k|y}$ and $ e^{-|k| y}$, 
the acceptable function for $y>0$ is $ e^{-|k| y}$ thus $\Phi(k,y)=\Phi(k,0) e^{-|k| y}$. 
The solution $\phi(x,y)$ being the inverse Fourier Transform of $\Phi(k,y)$ and since the latter is
the product of two Fourier Transforms  $\Phi(k,0)$ and  $e^{-|k| y}$, then by the convolution
theorem, $\phi(x,y)$  is the convolution of the inverse Fourier Transforms of 
$\Phi(k,0)$ and  $e^{-|k| y}$.

Using the boundary condition $\phi(x,y=0)=\phi_S(x)$ from eq.~\ref{BVP1} and using 
Fourier Transform definition~\ref{FT} results in:

\be
{\cal F}^{-1}[\Phi(k,0)]= {\cal F}^{-1} \left[\frac{1}{2\pi} \int_{-\infty}^{+\infty} \phi_S(x) e^{ikx} dx \right]=\phi_S(x)
\ee

Moreover recall that:
\be
{\cal F}^{-1}[e^{-|k| y}]=\frac{2y}{x^2+y^2} \equiv f(x,y)
\ee

since $k$ and $x$ are Fourier Transform pairs.

Applying the convolution theorem to the inverse Fourier Transforms yields the potential as:

\bea
\phi(x,y)&=&\frac{1}{2\pi} \int_{-\infty}^{+\infty} \phi_S(x') f(x-x',y) dx'\nonumber \\
         &=&\frac{1}{2\pi} \int_{-\infty}^{+\infty} \phi_S(x') \frac{2y}{{(x-x')}^2+y^2} dx'
\eea

which is exactly the solution eq.~\ref{2dgreen} found previously with 2D Green's function.

\section{Exact Fourier analysis}
Karlqvist solution of BVP1 eq.~\ref{BVP1} is now reexamined such that the gap field is no longer imposed.
Thus we have another BVP (called BVP2) corresponding to the exact solution given by:

\bea
\Delta \phi(\bm{r})&=&0 \hspace{1mm} \mbox{in Regions I and II}, \nonumber \\
\phi(\bm{r})&=&\pm V_0 \hspace{1mm} \mbox{on pole surfaces} \hspace{1cm} \mbox{(BVP2)}
\label{BVP2}
\eea 

The linear gap potential $\phi_{I}(x,y)$ found previously is corrected by adding a Fourier sum~\cite{Fan} 
with unknown discrete coefficients $A_n$ in order to satisfy BVP2 eq.~\ref{BVP2}. 
The potential $\phi_{II}(x,y)$ in the upper half-plane is expressed with continuous coefficients $C(k)$ obtaining:  

\begin{widetext}
\bea
\phi_{I}(x,y)=H_g x + \sum_{n=0}^\infty A_n \sin (n \pi x/a) e^{n \pi y/a} \hspace{2mm}  0 < x < a,  \hspace{2mm} y<0; \nonumber \\
\phi_{II}(x,y)=\int_0^\infty C(ka) \sin (kx) \hspace{1mm}  e^{-k y/a} dk \hspace{2mm} 0 < x < \infty, \hspace{2mm}  y>0
\label{Fouriersol}
\eea
\end{widetext}

In order to relate both sets of coefficients $A_n$ and $C(ka)$, we match the potential and derivative at $y=0$ obtaining:

\be
C(ka)= -\frac{I}{\pi} \frac{\sin(ka)}{{(ka)}^2} -  \sum_{n=1}^\infty {(-1)}^{n} 2n A_n \frac{\sin(ka)}{{(ka)}^2-{(n\pi)}^2}
\label{coeff}
\ee

\begin{figure}[htbp]
  \centering
    \includegraphics[angle=0,width=80mm,clip=]{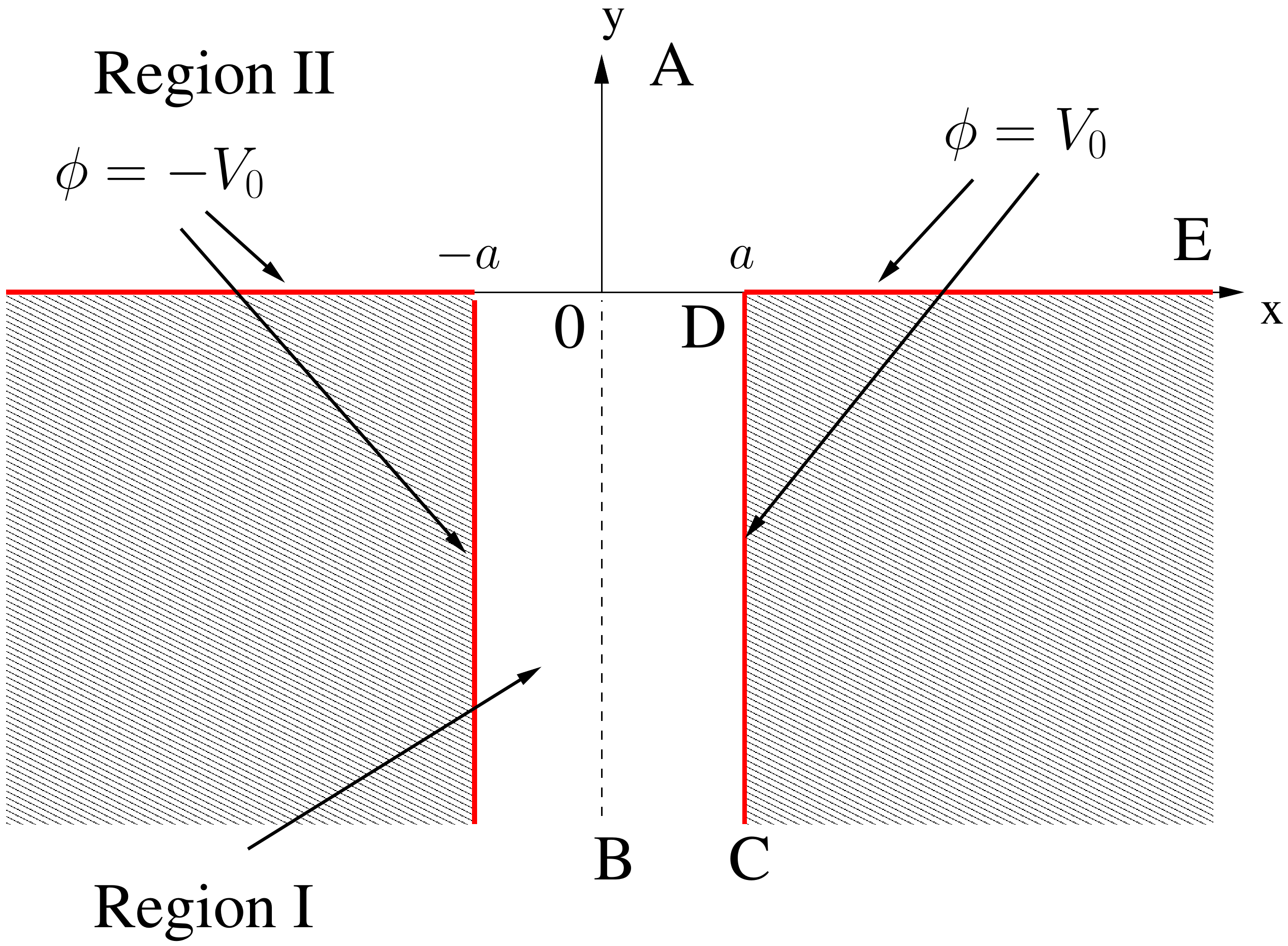}  
\caption{(Color on-line) Regions I and II where $\Delta \phi(\bm{r})=0 $ are bounded by red borderlines 
defining BVP2 eq.~\ref{BVP2} with pole surfaces considered as two distinct equipotentials $\phi= \pm V_0$.}
    \label{Fourier}
\end{figure} 

Eliminating coefficients $C(ka)$ allows us to derive relations among $A_m$ coefficients:

\be
\frac{A_m}{2}= \frac{2}{\pi} (-1)^{m+1}\left[\sum_{n=1}^\infty (-1)^{n} n \pi A_n I_{mn} + H_g I_{m0} \right]
\label{system}
\ee

The coefficients $I_{nm}$ are given in the Appendix. The above constitute
an infinite Algebraic system of linear equations that can be solved after numerical
truncation and performing LU decomposition techniques~\cite{Recipes} (see Appendix).
However it is preferable to use the exact values obtained from the combination of Fourier
results and conformal mapping described in the next section since the accuracy of the 
numerical values of $A_m$ decreases with the order $m$ when LU decomposition is performed.

\begin{figure}[htbp]
  \centering
    \includegraphics[angle=0,width=85mm,clip=]{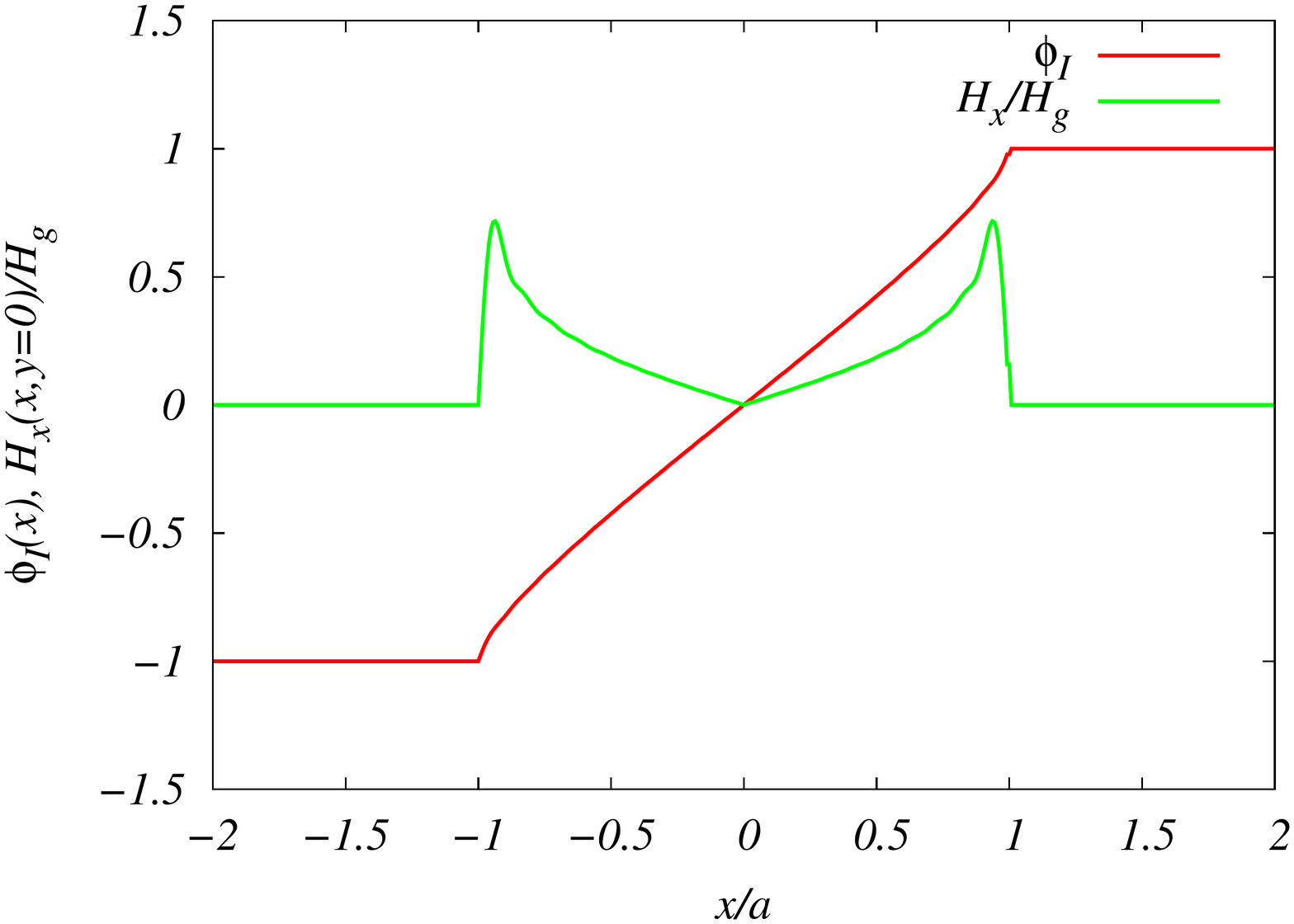}  
\caption{Fourier reconstructed potential $\phi_{I}(x,y=0)$  and field $H_x(x,y=0)$ in the gap.
Field oscillations due to Gibbs phenomenom arise from finite sum of 
Fourier coefficients. They are damped with Sigma Lanczos factors affecting as well the edge effect.}
    \label{gappot}
\end{figure}

\section{Conformal mapping} 
The Schwarz-Christoffel transformation~\cite{Ablowitz} is used to map the semi-infinite pole region 
of the $z$-plane onto the real  axis in the complex $w$ plane (see fig.~\ref{fig2}).

\begin{figure}[htbp]
  \centering
    \includegraphics[angle=0,height=85mm,clip=]{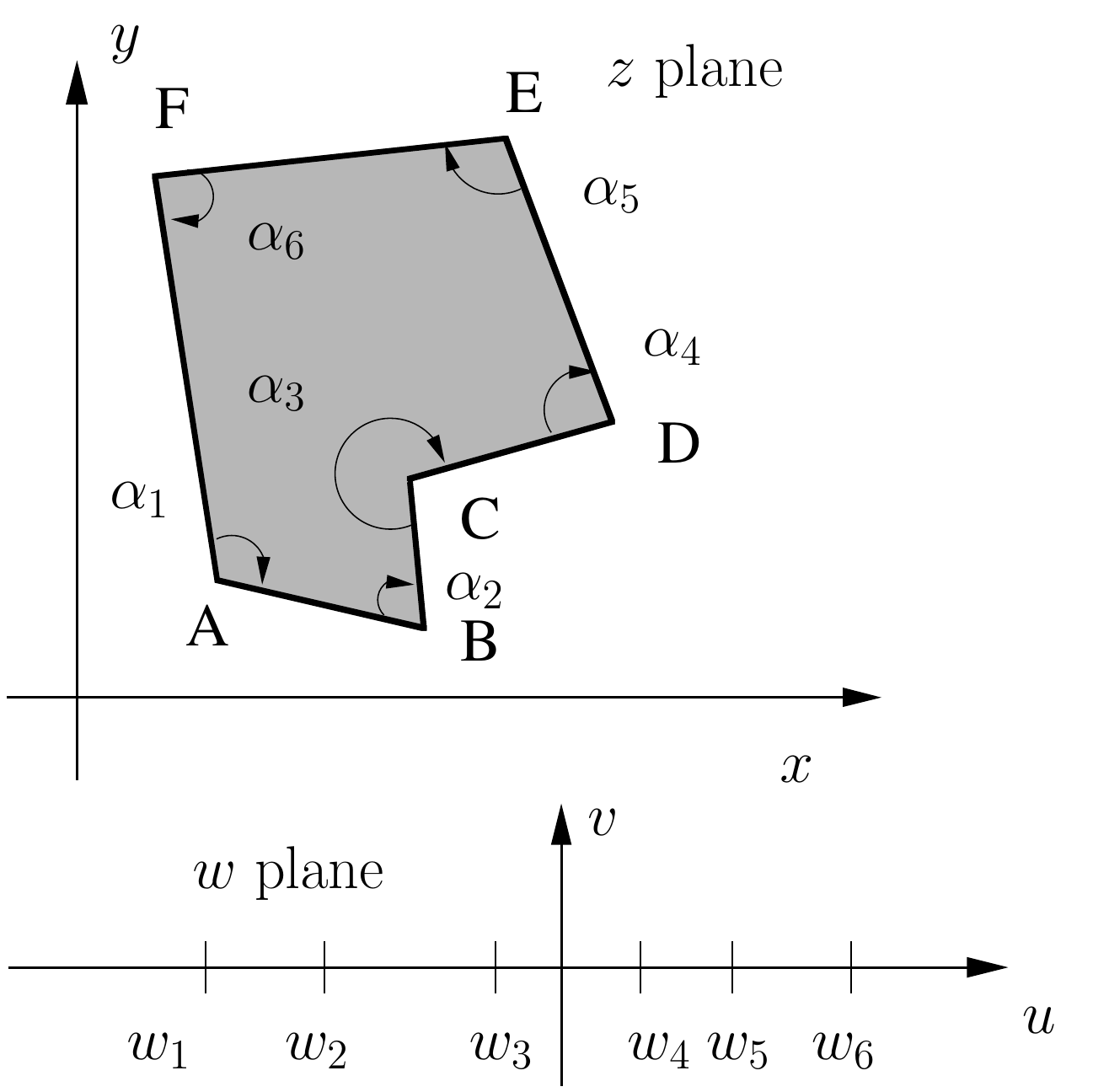}  
\caption{Schwarz-Christoffel mapping of the polygonal region ABCDEF
of the $z$-plane onto  $w_1,w_2,...w_6$ on the real axis of the complex $w$ plane.}
    \label{fig3}
\end{figure}

A polygon with vertices $a_i$ located in the $z$ plane with 
interior angles $\alpha_i$ is transformed into a sequence of points $w_1,w_2,...$
along the real axis in the $w$ plane with (see fig.~\ref{fig3}):

\be
\frac{dz}{dw}=\gamma {(w-w_1)}^{(\alpha_1/\pi)-1}  {(w-w_2)}^{(\alpha_2/\pi)-1}  {(w-w_3)}^{(\alpha_3/\pi)-1}...
\label{Schwarz0}
\ee

The inspection of polygon ABCDE in the $z$ plane (see fig.~\ref{fig2}) maps
BCD vertices with corresponding angles: $\alpha_1=\frac{\pi}{2},\alpha_2=\frac{\pi}{2},\alpha_3=\frac{3\pi}{2}$ 
into B'C'D' with abscissae $w_1=w_2=0, w_3=1$ with Schwarz-Christoffel relation:

\be
\frac{dz}{dw}=\gamma \frac{\sqrt{w-1}}{w}
\label{Schwarz}
\ee

where $\gamma$ is a complex constant to be determined.

Integration of eq.~\ref{Schwarz} yields:

\be
z=2 \gamma (\sqrt{w-1}-\tan^{-1} \sqrt{w-1}) +z_0
\label{map}
\ee

\begin{figure}[htbp]
  \centering
    \includegraphics[angle=0,width=80mm,clip=]{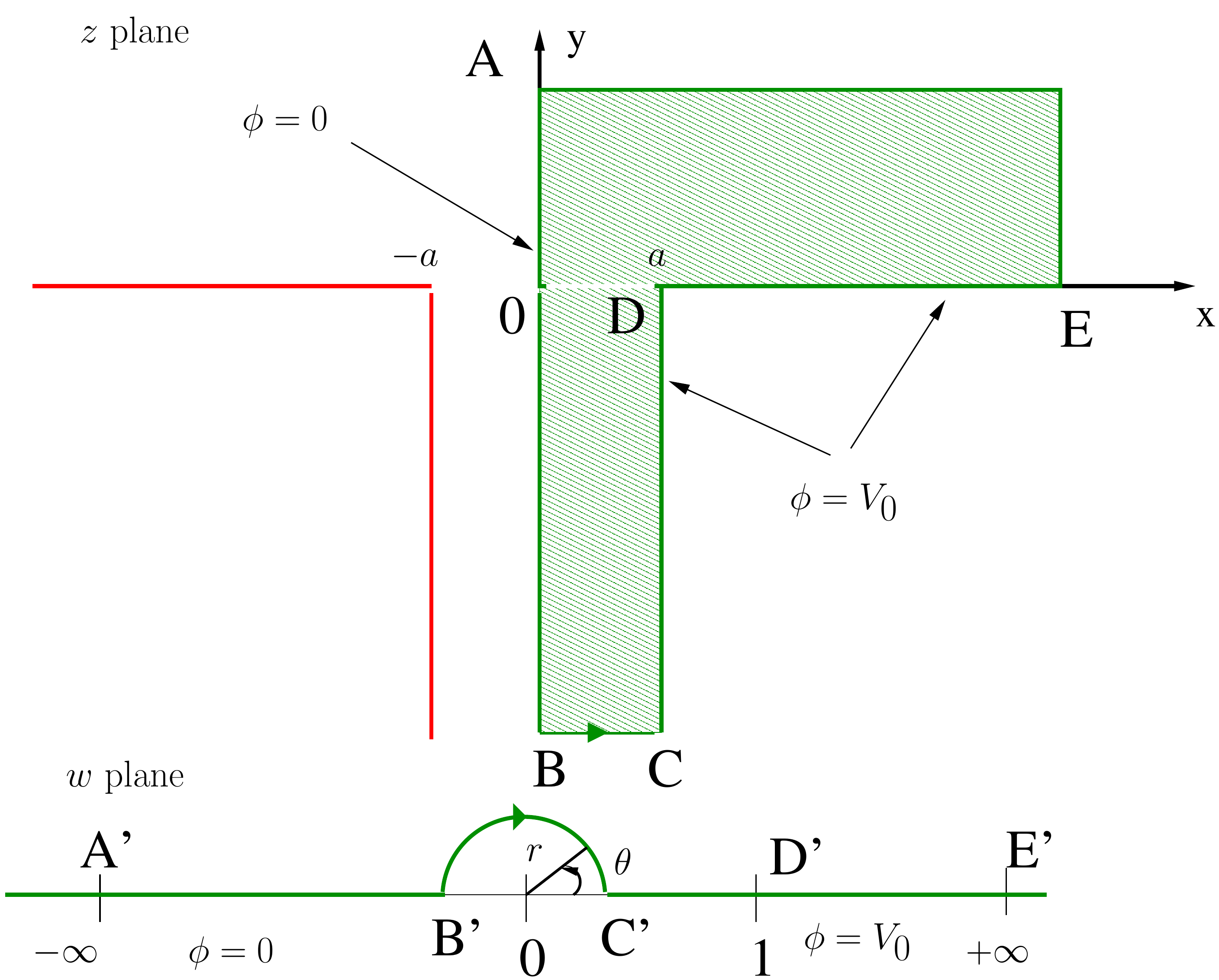}  
\caption{The polygonal region ABCDE in the $z$ plane is mapped onto A'B'C'D'E' along the real axis of the complex $w$ plane. Line BC is mapped to the line B'C' deformed into a half-circle with vanishing radius $r \rightarrow 0$  
to allow reveal the angle $\theta$  as 0 for C' and $\pi$ for B'. The potential $\phi=0$ over A'B' and $\phi=V_0$
over C'D'E'.}
    \label{complex}
\end{figure} 

Using $\tan^{-1} z= -i \tanh^{-1}(iz)$ and  $\tanh^{-1}(iz)= \frac{1}{2} \ln (\frac{1+z}{1-z})$ gives:

\be
z= 2\gamma \left[ \sqrt{w-1} +\frac{i}{2} \ln  \left( \frac{1+i\sqrt{w-1}}{1-i\sqrt{w-1}} \right) \right] +z_0
\label{Schwarz2}
\ee

In order to evaluate the unknowns $\gamma$ and $z_0$ we use the map eq.~\ref{map} D to D' 
thus $w=1$ corresponds to $z=a$ and we use the Schwarz-Christoffel transformation~\ref{Schwarz} 
to get:

\be
\int_B^C dz= \int_{B'}^{C'} \gamma \frac{\sqrt{w-1}}{w} dw
\label{angular}
\ee 

This contour integral is evaluated with the replacement $w=r e^{i\theta}$ with $r \rightarrow 0$
such that B' corresponds to $\theta=\pi$ and C' to $\theta=0$:

\be
a = \lim_{r \rightarrow 0} \int_{\pi}^{0} \gamma \frac{\sqrt{[r e^{i\theta}-1}]} {r e^{i\theta}} i r e^{i\theta} d\theta 
= \gamma \pi
\label{angular2}
\ee

Thus $\gamma =a/\pi$ and $z_0=a$.

After conformal mapping, we define an analytic complex function~\cite{Bertero} 
$F(w)=\psi+i\phi=-\frac{V_0}{\pi} \ln w + iV_0$
built with the magnetostatic flux $\psi$ and the potential $\phi$ satisfying the boundary
conditions in the $w$ plane, namely: $\phi=0$ along A'B' (since $\mbox{arg}(w)=\pi$),  
$\phi=V_0$ along C'D' and D'E' ($\mbox{arg}(w)=0$ )
(see fig.~\ref{fig2}). 

The magnetic field is obtained from $F(w)$ after differentiating with 
the complex operators $\frac{\partial}{\partial x}=(\frac{\partial}{\partial z}
+ \frac{\partial}{\partial \bar{z}})$ and $\frac{\partial}{\partial y}= i(\frac{\partial}{\partial z}
- \frac{\partial}{\partial \bar{z}})$ that transform field expression~(\ref{Laplace})
into a 2D complex form with $H_x=-\mbox{Im}\left[ \frac{dF(w)}{dz}\right]$ and 
$H_y=-\mbox{Re}\left[ \frac{dF(w)}{dz}\right]$. 

\begin{figure}[htbp]
  \centering
    \includegraphics[angle=0,width=85mm,clip=]{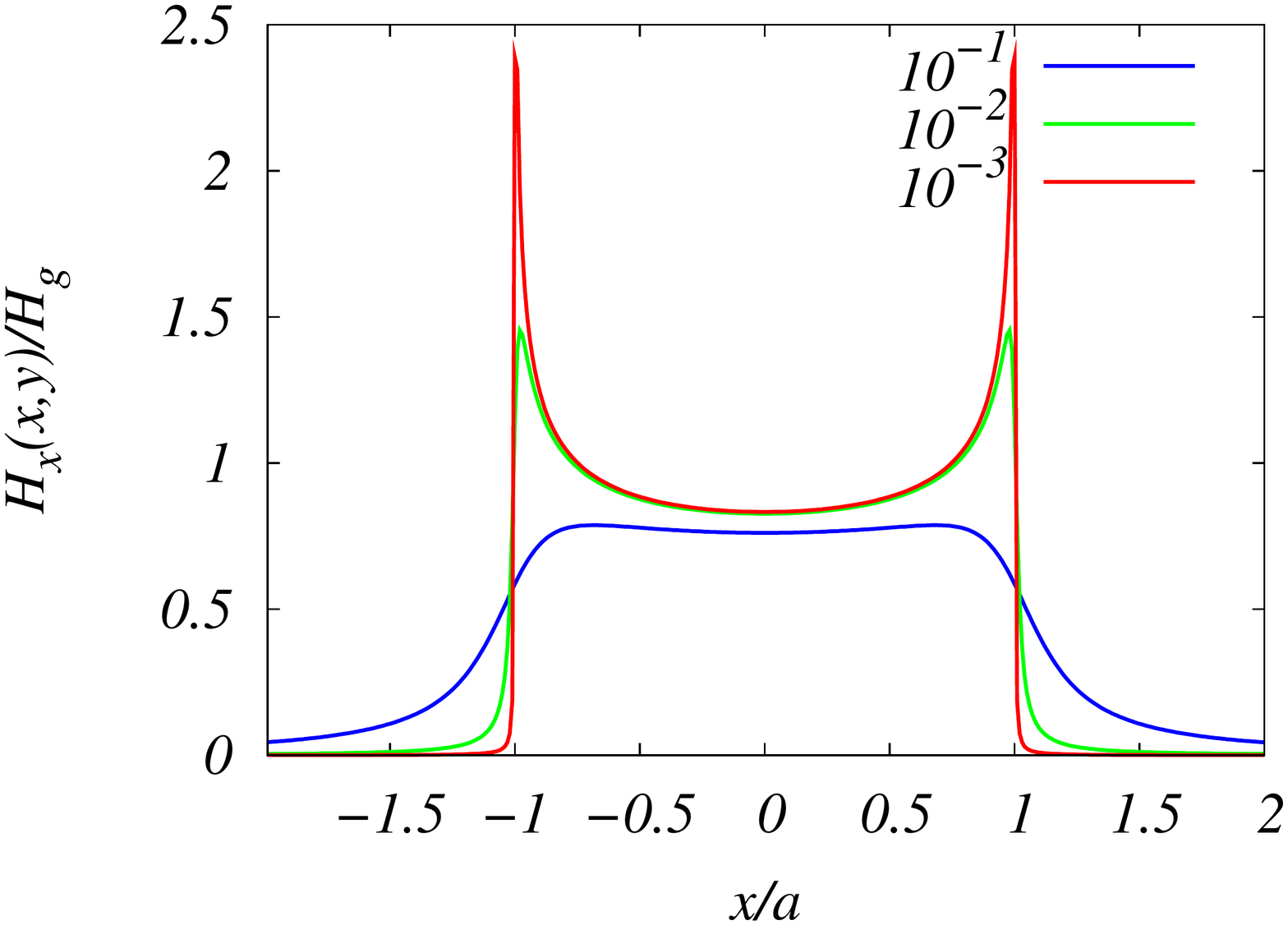} 
   \includegraphics[angle=0,width=85mm,clip=]{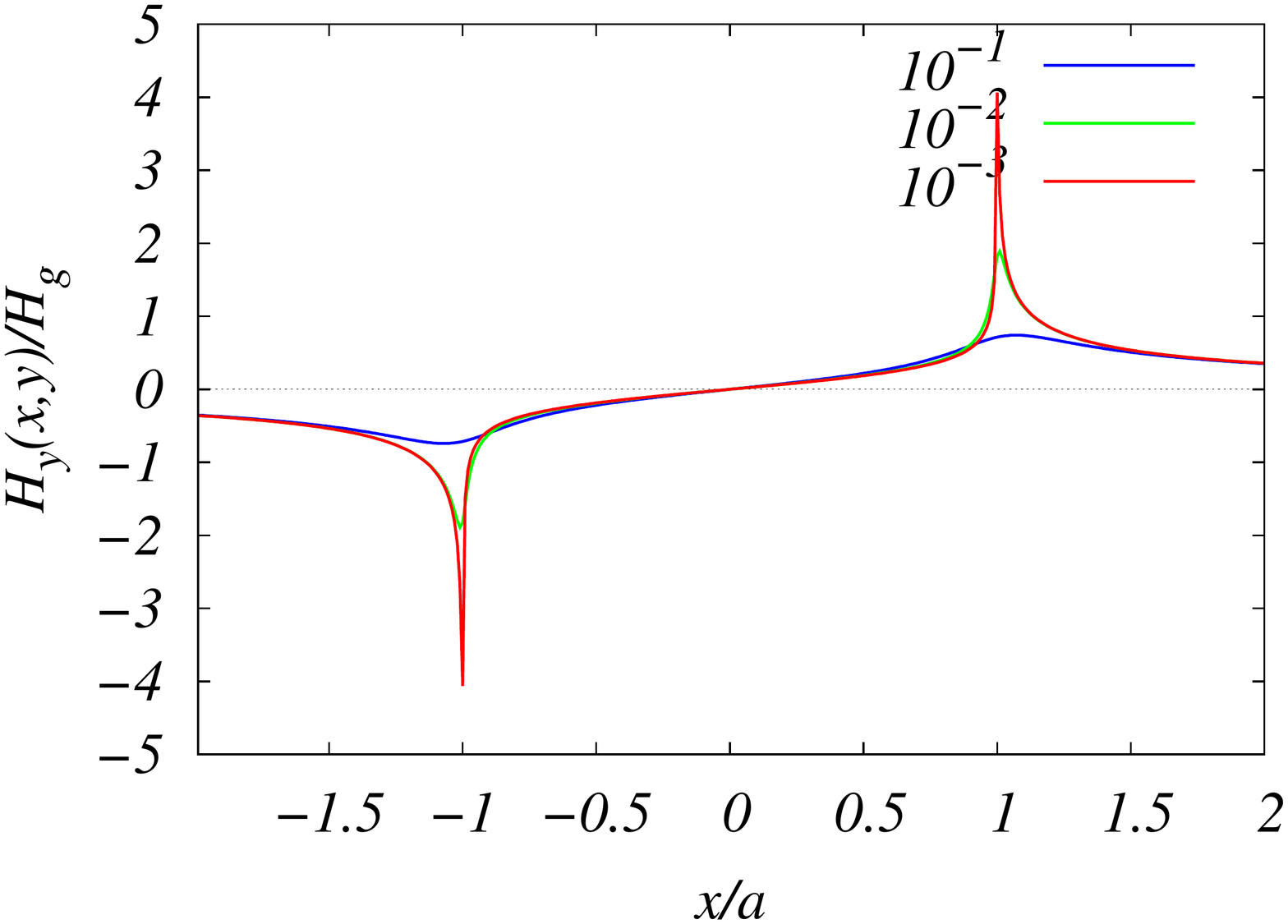}  
\caption{Variation of the fields $H_x(x,y), H_y(x,y)$ with $x/a$ for fixed $y/a=10^{-1},10^{-2},10^{-3}$.
Sommerfeld radiation singularities at $x=\pm a$  edges are enhanced as $y/a$ decreases.}
    \label{fig4}
\end{figure}

The fields $H_x(x,y), H_y(x,y)$ are obtained after relating $w$ to $z=x+iy$ using
relation~\ref{Schwarz2} and drawn versus $x/a$ for a fixed value of $y/a$ in fig.~\ref{fig4}.

$H_x(x,y)$ and $ H_y(x,y)$  behave around the edges according to the Sommerfeld radiation condition~\cite{Green}.
We find that $H_x, H_y \sim  y^{-\frac{1}{3} }$ for $x \approx \pm a$ which is expected from the
edge rule stating that the fields near an edge behave as $\frac{1}{\rho^{1-\nu} }$ with 
$\nu=\frac{\pi}{(2\pi -\alpha)}$ where $\rho$ is the distance to the edge and $\alpha$ the
edge angle. Applied to our case, we have $y=\rho, \alpha=\frac{\pi}{2}$ yielding $\nu=\frac{2}{3} $
thus the $y^{-\frac{1}{3} }$ behaviour confirmed by Matzner \etal~\cite{Matzner}.

\begin{figure}[htbp]
  \centering
    \includegraphics[angle=0,width=85mm,clip=]{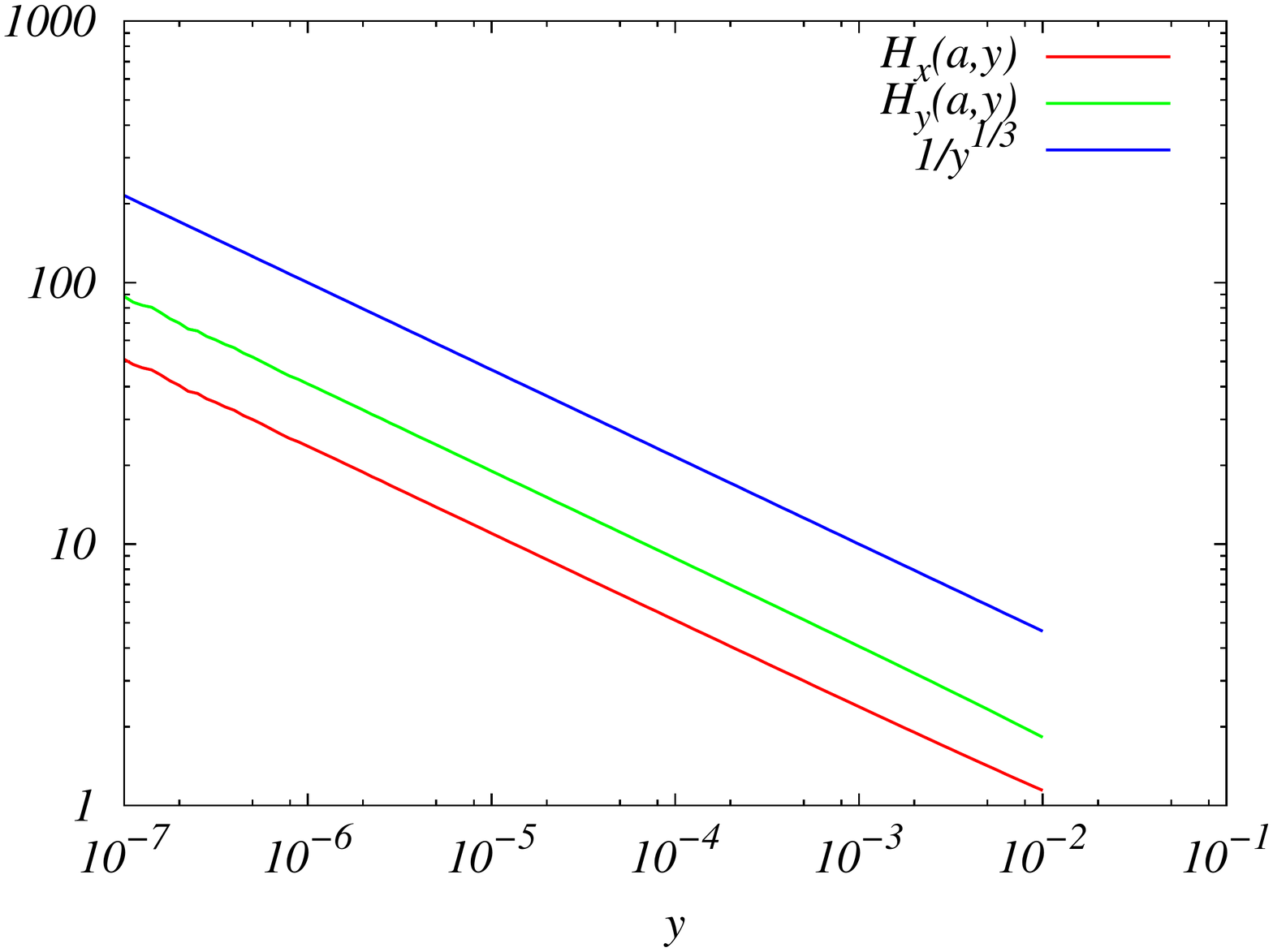}   
\caption{Scaling with $y$ of the edge fields $H_x(a,y), H_y(a,y)$ confirming the radiation condition
exponent predicted~\cite{Green} as 1/3.}
    \label{fig5}
\end{figure} 

\subsection{Exact Fourier $A_m$ coefficients}

Conformal mapping is now used to extract the exact~\cite{Wilton} $A_n$ coefficients that are used in
the Fourier solution eq.~\ref{Fouriersol}.

Since the gap solution given by $A_n$ coefficients that are related to the half-space 
solution coefficients given by  $C(ka)$, it might be useful to turn to Fourier transform
of the fields obtained from the conformal transformation.

The Fourier transform (cf. definition~\ref{FT}) of the head magnetic field ${\cal F}[{H}_x(x,y)]$
is related to the potential $\phi(x,y)$ Fourier transform  via $C(ka)$ such that~\cite{Wilton}:

\bea
{\cal F}[{\phi}(x,y)]=  -\frac{i\pi a}{2} C(ka) e^{ky}, \nonumber \\
{\cal F}[{H}_x(x,y)]=-\frac{k\pi a}{2} C(ka) e^{ky}.
\eea

Using $ka=m\pi$ in the $C(ka)$ expression~\ref{coeff} yields:

\be
A_m=  \frac{1}{m\pi} {\cal F}[{H}_x(x,y)] e^{m\pi y/a}
\ee

Taking $y=0$ gives $A_m=  \frac{1}{m\pi} {\cal F}[{H}_x(x,0)]$ and using relation
${\cal F}[{H}_x(x,0)]=V_0 S(ka)$ with $S$  given by Wilton \etal~\cite{Wilton}:

\be
S(m \pi)= \mbox{Im} \left[ \frac{1}{2\pi} \int_{-\infty}^{+\infty} \frac{s}{s^2+1}
\left(\frac{s+i}{s-i}\right)^m  e^{2mis} ds  \right],
\ee

it suffices to calculate the above integral in order to evaluate the exact $A_m$
coefficients.

The integral is straightforwardly calculated by complex contour integration~\cite{Wilton}. There is a single pole
$s=i$ of order $m+1$ in the upper half-plane. Integrating over the real axis and closing
the contour by a half-circle in the upper half-plane, we obtain from 
the $\frac{1}{(s-i)}$ coefficient of the $S(m \pi)$ Laurent series the value $K_m e^{-2m}$ with:

\be
K_m= \sum_{n=0}^{m-1} \binom {m-1} {n}  \frac{m+n}{(m-n)!} (-)^{m-n} (4m)^{m-n-1}
\ee

The coefficients $A_m= \frac{2 e^{-2m}}{m\pi}K_m$ are given in the Appendix and compared to Fourier results.

\section{Conclusion and perspectives}

The Karlqvist approximation pioneered study of magnetic recording and replay
heads. It is a very important landmark in the history of recording.

Karlqvist made a very elaborate approach based on a specific BVP (called BVP1) he
solved with Green's function techniques that were very popular at the time of his work.

While Karlqvist solution is based on a constant gap field, Fourier methods 
and conformal mapping methods tackle the Laplace problem directly as a
boundary value problem called BVP2 without considering the 
constant gap field assumption to obtain a more accurate and realistic
description of the head field satisfying Sommerfeld edge condition at angular points. 

While Green's function is important and yields several fundamental results, it lacks the
flexibility and versatility of conformal mapping that allows to derive the scaling
law at the edge points as well as evaluate the exact Fourier expansion coefficients.

It is remarkable that it took 40 years~\cite{Wilton90} to calculate the exact 
coefficients for a straightforward Laplace problem consisting of two 
semi-infinite equipotential domains separated by a gap and facing a free half-plane. 
Using popular software packages, be they symbolic (like Mathematica or Maple) 
and Finite-Element Method based such as COMSOL-Multiphysics can approach 
the problem readily with varying degree of success but miss the physical points
raised in this work. 

In spite of all developments in magnetic recording theory, 
Karlqvist approximation remains a major milestone in recording physics and 
is always used as a standard gauge with respect to any recording theory.

\appendix

\section{Evaluation of the exact Fourier Coefficients}

The numerical evaluation of the exact $A_n$ is straightforward but the 
evaluation of the infinite algebraic system of equations eq.~\ref{system} is tricky since it
involves non-trivial intermediate operations such as infinite summations,
evaluation of the $I_{mn}, I_{m0}$ integrals... 
In addition, the presence of minus signs in the system 
might induce numerical conditioning~\cite{Recipes}. 

Since the $A_m$ coefficients satisfy the relations (taking $H_g=1$ in eq.~\ref{system}):
\be
\frac{A_m}{2}= \frac{2}{\pi} (-1)^{m+1}\left[\sum_{n=1}^\infty (-1)^{n} n \pi A_n I_{mn} + I_{m0}\right]
\label{system1}
\ee

it is possible to write~\cite{Wilton90} for them a linear system
$M \bm{x}=\bm{b}$ where the matrix $M$ elements are given by:

\be
M_{mn}= (-1)^{m+n} n \pi I_{mn} + \left\{
\begin{array}{lcc}
 \frac{\pi}{4}  &  m=n, &  m=1,2,3... \\
 0  &  m \ne n, & n=1,2,3...
\end{array}
\right. 
\ee

the unknowns $x_m=A_m$ and the RHS terms are given by $b_m=(-1)^{m+1}I_{m0}, m=1,2,3...$.

The $I_{mn}$ terms are integrals given by: 

\be
I_{mn}=\int_0^{\infty} \frac{x \sin^2 x}{[{x^2- {(m \pi)}^2}] [{x^2- {(n \pi)}^2}]} dx
\ee

can be expressed after partial fraction decomposition into simpler integrals of the form:

\be
J_{m}=\int_0^{\infty} \frac{x \sin^2 x}{[{x^2- {(m \pi)}^2}]} dx=
 \frac{1}{2} \left[\int_0^{\infty} \frac{x}{[{x^2- {(m \pi)}^2}]} dx - L_m \right]
\ee

where:
\be
L_m=\int_0^{\infty} \frac{x \cos 2 x}{[{x^2- {(m \pi)}^2}]} dx
\ee

Integrals $L_m$  can be evaluated with contour integration~\cite{Ablowitz} or 
obtained from Gradshteyn-Ryzhik tables~\cite{Gradstein}.  
As a result, terms $I_{mn}, I_{mm}$ and $I_{m0}$ are expressed analytically as:
\begin{align}
I_{mn}&= \frac{[\ln(m/n)-Ci(2m\pi)+Ci(2n\pi)]}{2\pi^2(n^2-m^2)}  &  n \ne m, \nonumber \\
I_{mm}&= \frac{Si(2m\pi)}{2m\pi}  &  n=m,\nonumber  \\
I_{m0}&=-\frac{1}{2m^2 \pi^2}\left[ \gamma_E + \ln(2m\pi)-Ci(2m\pi)\right]  &  n=0 \nonumber
\end{align}

where $\gamma_E=0.57721566...$ is Euler-Mascheroni~\cite{Abramowitz} constant and $Ci(x),Si(x)$
are the cosine and sine integrals~\cite{Abramowitz} given respectively by:

\be
Ci(x)=-\int_x^\infty \frac{\cos(t)}{t} dt, \hspace{2mm} Si(x)=\int_0^x \frac{\sin(t)}{t} dt
\ee

In order to solve the system with LU decomposition~\cite{Recipes} we truncate it 
to an $N\times N$ system with $N$ taken successively as $N=100, 200...1000$. 
The convergence is monitored by comparing the LU coefficients to the exact coefficients
as $N$ varies. After establishment of convergence, a value of $N=1000$ allows
us to assess the accuracy in the evaluation of $A_m$ as $m$ increases (see Table~\ref{exact}).

\begin{table}[!h]
\begin{tabular}{l|c|c|c}
$n$ & Exact & LU decomposition & Wilton \etal~\cite{Wilton90} \\
\hline
1 & -8.6157113$ \times 10^{-2}$ & -8.6155161$ \times 10^{-2}$ & -8.6157121$ \times 10^{-2}$ \\
 2 &   2.9150246$ \times 10^{-2}$ & 2.9149190$ \times 10^{-2}$ & 2.9150244$ \times 10^{-2}$ \\
 3 & -1.5254218$ \times 10^{-2}$ & -1.5253429$ \times 10^{-2}$ & -1.5254219$ \times 10^{-2}$ \\
 4 &   9.5924977$ \times 10^{-3}$ & 9.5917108$ \times 10^{-3}$ & 9.5924996$ \times 10^{-3}$ \\
 5 & -6.6803270$ \times 10^{-3}$ & -6.6796015$ \times 10^{-3}$ & -6.6803293$ \times 10^{-3}$ \\
 6 &   4.9651614$ \times 10^{-3}$ & 4.9644802$ \times 10^{-3}$ & 4.9651619$ \times 10^{-3}$ \\
 7 & -3.8608240$ \times 10^{-3}$ & -3.8602557$ \times 10^{-3}$ & -3.8608965$ \times 10^{-3}$ \\
 8 &   3.1035715$ \times 10^{-3}$ & 3.1029664$ \times 10^{-3}$ & 3.1035778$ \times 10^{-3}$ \\
 9 & -2.5590272$ \times 10^{-3}$ & -2.5585052$ \times 10^{-3}$ & -2.5590907$ \times 10^{-3}$ \\
  10 &   2.1529666$ \times 10^{-3}$ & 2.1524425$ \times 10^{-3}$ & 2.1530113$ \times 10^{-3}$ \\
  11 & -1.8415294$ \times 10^{-3}$ & -1.8406087$ \times 10^{-3}$ & -1.8411567$ \times 10^{-3}$ \\
  12 &   1.5957345$ \times 10^{-3}$ & 1.5953275$ \times 10^{-3}$ & 1.5958599$ \times 10^{-3}$ \\
  13 & -1.3977018$ \times 10^{-3}$ & -1.3985117$ \times 10^{-3}$ & -1.3990281$ \times 10^{-3}$ \\
  14 &   1.2364594$ \times 10^{-3}$ & 1.2378943$ \times 10^{-3}$ & 1.2383976$ \times 10^{-3}$ \\
  15 & -1.1009253$ \times 10^{-3}$ & -1.1049084$ \times 10^{-3}$ & -1.1054002$ \times 10^{-3}$ \\
  16 &   9.9083269$ \times 10^{-4}$ & 9.9341292$ \times 10^{-4}$ & 9.9389255$ \times 10^{-4}$ \\
  17 & -8.9483510$ \times 10^{-4}$ & -8.9890108$ \times 10^{-4}$ & -8.9937093$ \times 10^{-4}$ \\
  18 &   8.2353526$ \times 10^{-4}$ & 8.1800920$ \times 10^{-4}$ & 8.1846816$ \times 10^{-4}$ \\
  19 & -7.4235769$ \times 10^{-4}$ & -7.4817217$ \times 10^{-4}$ & -7.4862316$ \times 10^{-4}$ \\
  20 &   7.0718257$ \times 10^{-4}$ & 6.8741635$ \times 10^{-4}$ & 6.8785821$ \times 10^{-4}$
\end{tabular}
\caption{Exact, LU decomposition and Wilton \etal evaluated $A_n$ coefficients~\cite{Wilton90}.
As $n$ increases, the discrepancy in the coefficients originates from decrease in accuracy of the LU decomposition.}
\label{exact}
\end{table}


\begin{thebibliography}{99} 
\bibitem{Karlqvist} O. Karlqvist, "Calculation of the Magnetic Field in The Ferromagnetic 
Layer of a Magnetic Drum", Trans. Roy. Inst. Techno., Stockholm, No. 86, 3 (1954).
\bibitem{Green} J. G. Van Bladel, {\it Singular Electromagnetic Fields and Sources},
Wiley-IEEE Press, New-York (1996). 
\bibitem{Ablowitz} M. Ablowitz and A. Fokas, {\it Complex variables: Introduction and Applications},
2nd edition, p. 287, Cambridge University Press, New-York (2003). 
\bibitem{Fan} G. J. Y. Fan, "A Study of The Playback Process of a Magnetic Ring Head", IBM J. of
Res. and Dev. 5, 321 (1961). Fan paper contains many typos that are corrected in: 
H.L. Huang and H. Y. Deng, "Comparison of Ring Head and SPT Head Write Fields",
IEEE Trans. Magn., Vol. 22, 1305 (1986).
\bibitem{Recipes} W. H. Press, W. T. Vetterling, S. A. Teukolsky and B. P. 
Flannery, {\it Numerical Recipes in C: The Art of Scientific Computing} Second Edition, 
Cambridge University Press, New-York (1992).
\bibitem{Bertero} G. A. Bertero, H. N. Bertram and  D. M. Barnett, "Fields and Transforms for Thin
Film Heads", IEEE Trans. Magn., Vol. 29, 67 (1993).
\bibitem{Matzner} H. Matzner and S. Shtrikman,"Some improved formulas for the Westmijze head", 
IEEE Trans. Magn., Vol. 33, 820 (1997).
\bibitem{Wilton} D. T. Wilton, B. K. Middleton and M. Aziz, "Exact Harmonic Coefficients for a
Magnetic Ring Head", IEEE Trans. Magn., Vol. 35, 2043 (1999).
\bibitem{Wilton90} D. T. Wilton, "Comparison of Ring and Pole Head magnetic fields",
 IEEE Trans. Magn., Vol. 26, 1229 (1990).
\bibitem{Gradstein}  I. S. Gradshteyn, I. M. Ryzhik, {\it Tables of Integrals, Series and Products}, 7th edition, Academic Press, New-York (2007) p.424, 3.723(11)
\bibitem{Abramowitz} M. Abramowitz and I.S Stegun, {\it Handbook of Mathematical Tables}, Dover, New-York (1960).
\end{thebibliography}
\end{document}